%% file: dilmtop_prd_v14.tex
\RequirePackage{lineno}

\documentclass[aps,prd,amsmath,amssymb,twocolumn,showpacs,superscriptaddress,groupedaddress]{revtex4}  % for review and submission
\usepackage{graphicx}% Include figure files
\usepackage{epsfig}
\usepackage{dcolumn}% Align table columns on decimal point
\usepackage{bm}% bold math
\usepackage{mathrsfs}

\makeatletter 
\renewcommand{\fnum@figure}{FIG.~\thefigure}
\makeatother

\newcommand{\ttbar}{t\bar{t}}

\newcommand{\met}{\mbox{$\protect \raisebox{.3ex}{$\not$}\et$}}
\newcommand{\MTreco}{M_{t}^{\rm reco}}
\newcommand{\MTalt}{M_{{\ell}b}^{\rm alt}}
\newcommand{\MTeff}{M^{\rm hyb}}
\newcommand{\Mtop}{M_{\rm top}}
\newcommand{\Mt}{M_t}
\newcommand{\et}{E_{T}}
\newcommand{\pt}{p_{T}}
\newcommand{\ppbar}{ p\bar{p} }
\begin{document}
\hspace{5.2in} \mbox{FERMILAB-PUB-15-190-E}
\setpagewiselinenumbers
%\linenumbers

% remove the following for publication
%\begin{figure}
%\leftline{\includegraphics[scale=0.5]{cdfii_thumb_logo.eps}\hfill
%/PUB/TOP/CDFR/11140}
%\end{figure}

% remove the space for publication
%\vspace*{1.75cm}

\title{Measurement of the top-quark mass in the $\ttbar$ dilepton channel \\ using the full CDF Run II data set}
%\author{The CDF Collaboration}
%\affiliation{URL http://www-cdf.fnal.gov}
%\date{\today}
\input{author}

\begin{abstract}
% remove the space for publication
%\vspace{1.3in}
\begin{center}
{\normalsize {\bf Abstract} }
\end{center}
We present a measurement of the top-quark mass in events containing two leptons (electrons or muons) with  a large 
transverse momentum, two or more energetic jets, and a transverse-momentum imbalance. 
We use the full proton-antiproton collision data set collected by the CDF experiment during the Fermilab Tevatron 
Run~II at center-of-mass energy $\sqrt{s} = 1.96$~TeV, corresponding to an integrated luminosity of 9.1~fb$^{-1}$. 
A special observable is exploited for an optimal reduction of the dominant systematic uncertainty, associated with the knowledge of
the absolute energy of the hadronic jets. The distribution of this observable in the selected events is compared to simulated distributions  of $\ttbar$ dilepton signal and background.We measure a value  for  the top-quark mass  of  
$171.5\pm 1.9~{\rm (stat)}\pm 2.5~{\rm (syst)}$~GeV/$c^2$.
\end{abstract}

% activate the following line for publication
\pacs{14.65.Ha, 13.85.Qk, 13.85.Ni}

\maketitle

\newpage
%===================== INTRODUCTION ==========================================================
\section{\label{sec:Intro}Introduction}
%\section*{\label{sec:Intro}}
In the standard model (SM) of particle physics, quark masses are proportional to their unknown Yukawa couplings to the Higgs field. 
Consequently, the masses are free parameters of the theory and must be determined experimentally.
Precise measurements of the top-quark mass ($\Mtop$) provide critical inputs to global fits of the electroweak parameters for checking the internal 
consistency of the SM~\cite{ewk} and for understanding the stability of the electroweak vacuum  at high energies~\cite{cite2}.

At the Fermilab Tevatron and the LHC colliders, measurements by the ATLAS, CDF,  CMS, and D0 Collaborations 
have given consistent results, 
whose combination has determined $\Mtop$ with a relative uncertainty of 0.44\%~\cite{mtop_comb}. 
The recent Tevatron combination  perfomed by  CDF  and D0   Collaborations  improved the relative incertainty to 
0.37\%~\cite{mtop_comb_tev}.    
All mass measurements in these  combinations 
were done analyzing events where the top quarks are produced  in pairs ($\ttbar$).
The top quark decays almost exclusively into a $W$ boson and a $b$ quark~\cite{pdg_l} and, depending on the
decay modes of the two resulting $W$ bosons, top quark-pair events yield final states with either 0, 1, or 2 charged leptons.
To improve the overall precision,  
the top-quark mass should be measured independently in all decay channels.
In the present analysis, we consider the events in the dilepton final state,
which is defined by the presence of two oppositely charged leptons (electrons or muons), 
two or more jets, and a large imbalance in the total transverse momentum  
from the two neutrinos associated with the charged leptons (``$\ttbar$ dilepton events'' 
or ``dilepton channel'').

At the Tevatron, the most accurate  $\Mtop$ measurements in the dilepton channel~\cite{Hyun-Su5.6,D0-dil2}
%{,cms_dil, Atlas_dil2} 
use methods 
%{Hyun-Su5.6,D0-dil,cms_dil,atlas_dil} 
of full or partial  reconstruction of the top-quark events.  In these analyses, the systematic  uncertainty 
dominates over the  statistical one with a large contribution of the 
jet-energy scale (JES) uncertainty.
Measurements in the other final states reduce the JES systematic uncertainty by constraining the mass of the final-state 
jet pair to match the $W$-boson mass. This constraint permits a precise calibration of the calorimeter JES~\cite{mmm1, mmm2, mmm3}.
Since dilepton $\ttbar$ events do not contain jets from $W$ decays, we devise a new method 
to reduce the impact of the JES uncertainty on the measurement result. In the  past, CDF developed two 
methods to  reconstruct the top-quark mass  using  only quantities with minimal dependence on the JES.
One measurement exploited the transverse decay length of $b$-tagged jets~\cite{Lxy}
and another the transverse momentum of electrons and muons from $W$-boson decays to determine the 
top-quark mass~\cite{Lepton_Pt,Lxy}.
These  methods decreased  the  systematic  uncertainty stemming from the JES uncertainty, but suffered from an increase of the 
statistical uncertainty due to  their  low sensitivity  to the top-quark mass.  
In the current  analysis,  we  combine two  reconstruction methods, one with a strong dependence 
and one  with a  minimal dependence on JES.  The combined method  
simultaneously  optimizes the effect of the statistical and systematic  uncertainties delivering a  result with  
a  minimal  total  uncertainty.

This paper reports on the  final CDF  $\Mtop$ measurement in the dilepton channel 
performed with proton-antiproton collision data at $\sqrt{s}=1.96$~TeV, 
collected at the Tevatron with the CDF~II detector~\cite{cdf_l}. 
The measurement uses the full CDF Run~II data set accumulated between March 2002 and September 2011
and  corresponds to an integrated luminosity of 
9.1~fb$^{-1}$. The results supersede those of Ref.~\cite{Hyun-Su5.6}  by exploiting an improved analysis technique 
and 
%in the dilepton final state by exploiting 
an  additional integrated luminosity of about 3~fb$^{-1}$. 

%The measurement corresponds to an integrated luminosity of 
%9.1~fb$^{-1}$. It uses the full CDF Run~II data set accumulated between March 2002 and September 2011
%and  supersedes the previous CDF result~\cite{Hyun-Su5.6} in the dilepton final state by exploiting 
%an  additional integrated luminosity of about 3~fb$^{-1}$. 

%The method applied to the increased data sample 
%results in reducing the  total uncertainty by 14\% with respect to the previous measurement~\cite{Hyun-Su5.6}, 
%where 9\% is coming from the method itself. 

%%%%%%%%%%%%%%%%%%%%%%%%%%%%%%%%%%%%%%%%%%%%%%%%
%===================== DATA SAMPLE AND EVENT SELECTION =======================================
\section{\label{sec:Data}Detector, Data Sample and Event Selection}
%\section{\label{sec:Det}CDF II Detector}

The CDF~II  detector is a general-purpose apparatus~\cite{cdf_l} designed to detect the products of $\ppbar$ collisions at the Tevatron.
It consists of a magnetic spectrometer surrounded by calorimeters and muon detectors.
The spectrometer has a charged-particle tracking system consisting of a silicon microstrip tracker and a drift chamber.
The tracking system is immersed in the 1.4~T magnetic field of a solenoid aligned with the beams.
Segmented  towers of electromagnetic and hadronic sampling calorimeters,
located outside the solenoid, measure particle energies.
A set of drift chambers and scintillation counters surrounds the calorimeters and detects muons.
The detector has an approximate cylindrical geometry around the Tevatron beamline, which  
makes   convenient to use a  cylindrical  coordinate system~\cite{csyst}  to 
describe the kinematic properties of reconstructed events.

%\section{\label{sec:Data}Data Sample and Event Selection}
The data are collected with an inclusive online event selection (trigger) that requires  
an electron (or a muon) with transverse energy $\et>18$~GeV (transverse momentum $\pt>18$~GeV/$c$) 
in the central pseudorapidity region ($|\eta| < 1.1$) of the detector.
Offline, the sample is further selected using the criteria developed for the $\ttbar$ 
cross section measurement in the dilepton channel~\cite{xsec8.8}. 
In this analysis we introduce additional requirements to improve event modeling and to reduce 
the total background.

For the selection of events
we require the presence of two oppositely charged leptons  ($\ell$), with $\et>20$~GeV for electrons or $\pt>20$~GeV/$c$ for muons, 
at least one of which must be isolated~\cite{isol} and detected in the central region of the detector ($|\eta| < 1.1$).
We further require large missing transverse energy~\cite{met_def}, $\met>25$~GeV,
and at least two jets with $\et>15$~GeV and $|\eta|<2.5$.
To detect jets we look for clusters of energy in the calorimeter 
using a cone algorithm with radius $R=\sqrt{(\Delta \eta)^2+(\Delta \phi)^2}=0.4$~\cite{ConeAlg},
where $\phi$ denotes the azimuthal angle.
Jet energies are corrected for instrumental effects~\cite{JetCorrections}.
In events in which $\met$ originates from mismeasurements of the leptons or jets, the   azimuthal angle between
the $\met$ vector and the direction of the mismeasured object is typically small.
To suppress this instrumental background, we increase the $\met$ requirement to $\met>50$~GeV
for events where $\Delta\phi$ between the  directions of $\met$ and at least one of the reconstructed leptons or jets 
is less  than $20^\circ$. 
%relative to the directions of jets or leptons.
One of the main backgrounds is due to events in which a $Z$ boson is produced in association
with jets and  decays to an electron or muon pair ($Z\rightarrow ee, \mu\mu+{\rm jets}$). These  events  
may feature large $\met$ due to a mismeasurement of the leptons or jets.
Therefore, a supplementary requirement is applied to $e^+e^-$ and $\mu^+\mu^-$ events 
when the dilepton mass is within 15~GeV/$c^2$ of the known $Z$-boson mass~\cite{pdg_l}.
For these events, we require a $\met$  significance~\cite{met_sig} in
excess of $4\mbox{~GeV}^{1/2}$.
Since the products of  $\ttbar$ decays have  large transverse energies, 
further background suppression is achieved by requiring $H_T > 200$~GeV~\cite{ht_def}.   
Another large source of background is due to events in which a $W$ boson produced in association with
jets ($W$+jets) yields a single lepton in the final state, 
where one of the jets is misidentified as a second lepton (``$W$+jets fakes''). 
We find that approximately half of these events  
feature a small distance in the $\eta$-$\phi$ space 
between the fake lepton  and the axis of one of the jets ($j$), 
$\Delta R_{{\ell}j}=\sqrt{(\Delta \eta_{{\ell}j})^2+(\Delta \phi_{{\ell}j})^2}$.
To reject this background we require $\Delta R_{{\ell}j}$ to be greater than 0.2 
for all possible pairings between leptons and jets in the event.

To obtain  the  most probable  value of the top-quark mass per event ($\MTreco$), we  use  
a kinematic reconstruction method. This  method calculates $\MTreco$ using all  of 
the available experimental event information 
and has optimal sensitivity to $\Mtop$. 
From simulation, 6\% of background events have $\MTreco$ 
larger than 250~GeV/$c^2$, while only about 0.5\% of signal, simulated with $\Mtop$ between $160$~GeV/$c^2$ to $185$~GeV/$c^2$,
contributes to this region.
In the  analysis, we reject  the  events with  $\MTreco>250$~{\rm GeV}/$c^2$.
Finally, the dilepton invariant mass is required to be larger than 10~GeV/$c^2$ 
to suppress events from the decays of low-mass dimuon resonances and to improve the background modeling.
In total we have 520 $\ttbar$ dilepton candidates that pass the selection requirements.

The sensitivity of the  measurement is improved by analyzing separately events with a jet identified
  as originating from the fragmentation a bottom quark ($b$-tagged).
We divide the event sample into two independent subsamples.
The first subsample ($b$-tagged sample) contains events with at least one $b$-jet tagged using the secondary vertex 
({\sc secvtx}) $b$-tagging algorithm~\cite{b_tag_alg}. This  algorithm uses 
information from the displacement of secondary vertices relative to the primary event vertices to 
``tag'' $b$-hadron decays. The second subsample contains events 
in which no $b$-tag is found (non-tagged sample).

The {\sc pythia}~\cite{Pythia} Monte Carlo (MC) program with CTEQ5L~\cite{CTEQ5L} parton distribution functions is used to generate
samples of $\ttbar$ events with various top-quark masses.
All MC samples are generated in combination with a detailed simulation of the CDF~II detector~\cite{CDFIISimulation}.
Depending on the process, backgrounds are modeled using simulated or experimental data.
The MC samples of diboson events (${\it WW}$, ${\it WZ}$, and ${\it ZZ}$) are obtained 
using {\sc pythia} whereas
the Drell-Yan events ($Z/\gamma^*$+jets, $Z/\gamma^*\rightarrow ee,\mu\mu,\tau\tau$)  are generated with the {\sc alpgen} program~\cite{ALPGEN} interfaced to {\sc pythia} for 
showering and hadronization. A detailed description of the CDF MC procedures and samples
 is provided in Ref.~\cite{Hyun-Su1.9}.
The background originating from events in which a jet is misidentified as a lepton 
is modeled with $W$+jets data.
The composition of the data sample is estimated using the methods described in Ref.~\cite{xsec8.8}.
Table~\ref{tab:sample_tab_000} 
summarizes the expected and observed $\ttbar$ signal and background yields. The 
signal yield is calculated assuming 7.4~pb for the $\ttbar$  production cross-section.
The Drell-Yan and $W$+jets (``fake'') events are  
the main sources of contamination.  % in the final $\ttbar$ dilepton sample.The comparison presented in 
Table~\ref{tab:sample_tab_000}  shows  excellent agreement between expected and observed event yields.
\begin{table}
%\parbox{.45\linewidth}{
%\begin{footnotesize}
%\begint{flushleft}
\caption{Number of expected and observed events in the $b$-tagged and non-tagged samples.~~~~~~~~~~~}
\label{tab:sample_tab_000}
\begin{center}
%\caption{Number of expected and observed events in the $b$-tagged and non-tagged samples.}
\vspace{0.05in}
\begin{tabular}{lcc}
\hline
\hline
% \multicolumn{3}{|c|}{ \bf $\ttbar$ dilepton sample } \\
%\cline{1-3}
%    Source         &  Tagged events    & 0 tag events               \\
    Source         &  $b$-tagged           & Non-tagged                \\
                   &  sample               & sample    \\
\hline
 $WW$                 & 0.6 $\pm$ 0.2  &   16.4 $\pm$ 3.6        \\
 $WZ$                 & 0.1 $\pm$ 0.0  &    5.2 $\pm$ 1.0        \\
 $ZZ$                 & 0.2 $\pm$ 0.1  &    3.0 $\pm$ 0.5       \\
 Drell-Yan            & 4.4 $\pm$ 0.4  &   51.2 $\pm$ 8.0       \\
 Fakes                & 8.6 $\pm$ 2.7  &   21.4 $\pm$ 6.2  \\
\hline
Total background         & 13.9 $\pm$ 2.8   &   97.2 $\pm$ 14.5    \\
$\ttbar$ ($\sigma=7.4$~pb)   & 227.2 $\pm$ 16.2  & 173.2 $\pm$ 13.3          \\
\hline
 Total SM expectation        &  241.1 $\pm$ 16.4   &  270.3 $\pm$ 26.4  \\
\hline
 Observed      & 230 & 290   \\
\hline
\hline
\end{tabular}
%\vspace{0.04in}
%\caption{
%Summary table with the number of expected and observed events in the $b$-tagged and non-tagged data samples.
%}
%\label{tab:sample_tab_000}
\end{center}
%\end{footnotesize}
\end{table}

%===================== Calculating variable for Top Quark Mass Measurement ========================================
\section{\label{sec:Method} Methodology}
The template technique~\cite{template} used in this analysis estimates the top-quark mass by performing
a fit of the distribution of an observable to a sum of signal and background contributions.
This method can be applied to any observable whose distribution depends on $\Mtop$. 
However, the choice of the observable has direct impact on the precision of the measurement. 
For this analysis, we develop a variable  that is expected to achieve a minimal measurement uncertainty.
We start from two initial observables: the first observable is $\MTreco$, which 
is computed using the ``neutrino $\phi$-weighting method''~\cite{PreviousAnalysisArt}.
%This method was previously used by CDF for a top-quark mass reconstruction in the analysis  described in Ref.~\cite{PreviousAnalysisArt}.
To account for the unconstrained kinematics of the top-quark decay, we scan over the phase space of the azimuthal angles 
of  both neutrino momenta  and for each point of this two-dimensional scan we  reconstruct the top-quark mass by minimizing 
a $\chi^2$ function for the $\ttbar$ final state hypothesis. Following the  scan, 
we assign $\chi^2$-dependent weights to the solutions in order to identify a preferred  $\MTreco$ for each event.
Since  this  method  uses all of the  event information, including  the  jet energies,
the  reconstructed   mass  strongly depends on the calorimeter JES.

To reduce this systematic dependence, we consider a second observable that is insensitive to the JES.
Testing a number of observables defined without using any information about jet energies, we choose the one that 
has the best sensitivity to $\Mtop$. 
This observable, denoted as ``alternative'' mass ($\MTalt$), is defined according to the following formula:
\begin{eqnarray}\label{var2_2}
\MTalt &=& c^2 \sqrt{ \frac{\langle {\ell}_1,b_1 \rangle \cdot \langle {\ell}_2, b_2 \rangle}{E_{b_1} \cdot E_{b_2}}},
\end{eqnarray}
where $\ell{_1}$ and $\ell{_2}$ are the four-momenta of leptons, and $b_1$ and $b_2$ are 
the four-momenta of the two highest-$\et$ (``leading'') jets,
which are defined as for massless particles, with energies $E_{b_1}$ and $E_{b_2}$.
The quantity $\langle l$,$b \rangle$ indicates the scalar product of the ${\ell}$ and $b$ four-vectors.
The jet energies $E_{b_1}$ and $E_{b_2}$ appear in the denominator of Eq.~(\ref{var2_2}) to  cancel 
the $\MTalt$ JES-dependence of the leading jets, present  in the numerator. 

The use of the two leading jets in Eq.~(\ref{var2_2}) is justified because 
in about 78\% of the selected $\ttbar$ events the two leading jets originate from the hadronization of the two $b$ quarks 
in the $\ttbar$ decay, according to simulation.
We use the same index (1 or 2) to indicate a lepton and a jet that are assumed to originate 
from the decay of the same top quark.
To choose between the two possible pairings of leptons and $b$-jets, we select the configuration 
%with the closest  proximity between the  a leptons and a jets  directions for the  two pairs.  
%For  this  selection we use 
with  the  maximum value of the scalar products  
 $\langle \boldsymbol{c}_{l_1},\boldsymbol{c}_{b_1} \rangle + \langle \boldsymbol{c}_{l_2},\boldsymbol{c}_{b_2} \rangle$
where  $\boldsymbol{c}$  is  an unit vector collinear with the  lepton or $b$-jet directions and the 
indexes $l_1$ and $b_1$ ($l_2$ and $b_2$) correspond to the lepton and $b$-jet in the first (second) pair.  
%, the proximity of lepton directions  and jets is examined.
%To choose between the two possible pairings of leptons to jets, the proximity of lepton directions  and jets is examined.
%We specify the track direction of the leptons and $b$ jets through the unit vector $\boldsymbol{c} = (c_x, c_y, c_z)$,
%where $c_x$, $c_y$, and $c_z$ are 
%the direction cosines of their momenta~\cite{dir_cos}.
%The pairing with the largest sum of scalar products
%$\langle \boldsymbol{c}_{l_1},\boldsymbol{c}_{b_1} \rangle + \langle \boldsymbol{c}_{l_2},\boldsymbol{c}_{b_2} \rangle$
%is chosen.
%The indexes $l_1$ and $b_1$ ($l_2$ and $b_2$) correspond to the lepton and $b$-jet in the first (second) pair.
From simulation, we estimate that this lepton-to-jet pairing criterion selects the right pairing in  $61\pm1\%$ of the cases.
Other pairing criteria provide higher pairing efficiency of about 70\%. 
However, these criteria  use JES-dependent variables that create undesirable correlations between $\MTalt$ and $\MTreco$.   
% spoiled the insensitivity of $\MTalt$ to the JES because they are JES-dependent. 

We define the ``hybrid'' variable $\MTeff$,
\begin{eqnarray}
\label{hybrid}
\MTeff = w \cdot \MTreco + (1-w) \cdot \MTalt,
\end{eqnarray}
where $w$ is a weighting parameter between 0 and 1.
The statistical and systematic uncertainties of the measurement depend on the choice of the $w$ parameter.
{\it A priori}, we choose the value of $w$  that gives the smallest combined statistical and systematic uncertainty based on simulation.
In order to find the optimal value of $w$, we scan the [0,1] interval in steps of 0.05.
For every point of the scan, we define the mass fit  using the signal and background templates for $\MTeff$
and evaluate the uncertainties.

Signal templates for $\MTeff$ are formed separately for $b$-tagged and non-tagged events from $\ttbar$ samples generated for top-quark masses $\Mtop$ 
in the range from $160$~GeV/$c^2$ to $185$~GeV/$c^2$ with a 1~GeV/$c^2$ step. 
The probability density functions (p.d.f.'s) of the signal, which 
express the probability of getting any $\MTeff$ value in $\ttbar$ events 
with given $\Mtop$,
are obtained as parametrizations of the corresponding templates.
We parametrize the templates using a sum of two Landau and one Gaussian probability distribution functions. 
The parameters of these p.d.f.'s depend linearly  on $\Mtop$. 
The background templates are derived separately for $b$-tagged and non-tagged events by adding diboson, fake, and Drell-Yan templates 
that are normalized to the expected rates reported in Table~\ref{tab:sample_tab_000}. 
The background p.d.f.'s are obtained from a likelihood fit of the combined background templates, 
performed in the same way as  for the signal templates, but without any dependence on the top-quark mass.

%\section{\label{sec:LHFit} Likelihood Method}
To measure $\Mtop$ we perform 
a likelihood fit of the unbinned data distributions 
to a weighted sum of signal and background p.d.f.'s.
The mass returned by the fit corresponds to the maximum of a likelihood function ($\mathscr{L}^{\rm total}$)
defined as the product of independent likelihood functions obtained 
for $b$-tagged and non-tagged subsamples $\mathscr{L}^{\rm total}= \mathscr{L}^{\rm tag} \cdot \mathscr{L}^{\rm non-tag}$.
The terms $\mathscr{L}^{\rm tag}$ and $\mathscr{L}^{\rm non-tag}$ represent 
the probabilities 
%for the corresponding subsamples 
that the  $\MTeff$ distribution observed in data  
comes from a mixture of background events and $\ttbar$ dilepton events with an assumed top-quark mass $\Mtop$. 
The  $\mathscr{L}^{\rm tag}$ and $\mathscr{L}^{\rm non-tag}$ form is similar to the likelihood function 
used in Refs.~\cite{an2005, PreviousAnalysisArt} and can be written as
\begin{eqnarray}
\begin{split}
\mathscr{L}^{i} =
\mathscr{L}^{\rm bg}_{\rm constr}(n_b^i) \cdot
\mathscr{L}_{\rm stat}(N^i|n_s^i+n_b^i) \cdot \\
\prod_{k=1}^{N^i}
\mathscr{L}_{\rm evt}^k(\MTeff{^{,\; k}}|\Mtop,n_s^i,n_b^i),
\end{split}
\label{lhood0000}
\end{eqnarray}
where $N^i$ is the number of events  in the corresponding subsample $i$.
Using the signal and background p.d.f's, the likelihood term $\mathscr{L}^{k}_{\rm evt}$ represents  the probability for an event $k$
with mass $\MTeff{^{,\; k}}$
to be observed in sample $i$ where $n_s^i$ and $n_b^i$ events are expected for signal and background, respectively.  
The term $\mathscr{L}_{\rm stat}$ gives 
the probability of observing $N^i$ events in the sample, according to a Poisson distribution, while
$\mathscr{L}^{\rm bg}_{\rm constr}$ constrains the number of background events in the corresponding subsample to the value shown 
in Table~\ref{tab:sample_tab_000}.
Having as inputs the $\MTeff$ values observed in  data, the signal and background p.d.f.'s, and the expected background, 
the likelihood fit returns 
the estimated top-quark mass ($\Mt^{\rm fit}$) and the estimated number of signal and background events.
% ($n_s^{\rm tag,\; fit}$, $n_s^{\rm 0tag,\; fit}$, $n_b^{\rm tag,\; fit}$ and $n_b^{\rm 0tag,\; fit}$).

Since $\MTeff$ depends on $w$, the likelihood fit is different at each point of the $w$-scan. 
The correctness of these $w$-dependent fits is checked with simulated experiments (``pseudoexperiments'' or PE's) 
performed on samples of MC events with given input top-quark mass ($\Mtop^{\rm inp}$).
%The PE's confirm  that $M_{t}^{\rm fit}$ is an unbiased estimate of $\Mtop$, and that its uncertainty is correctly estimated.
In every PE we draw the number of signal and background events according to Poisson distributions with means given in Table~\ref{tab:sample_tab_000} and then draw values of $\MTeff$ according to the corresponding signal and background templates. PE's obtained in this way are used in our check of likelihood fitting. They  
confirm  that $M_{t}^{\rm fit}$ is an unbiased estimate of $\Mtop$ and  its uncertainty is also correctly estimated.
%Performing for each $\Mtop^{\rm inp}$ a number of PE's  we check that
%$M_{t}^{\rm fit}$ is an unbiased estimate of $\Mtop$, and that its uncertainty is correctly estimated.

In order to choose which $w$-dependent likelihood fit is to be applied to the data, we estimate
the uncertainties as  functions of $w$.
We define the expected statistical uncertainty as the average statistical uncertainty 
in PE's with $\Mtop^{\rm inp}=172.5$~GeV/$c^2$.
To evaluate the JES systematic uncertainty, we test  the impact 
of the uncertainties associated with the following effects:
non-uniformity in calorimeter response as a function of $|\eta|$, 
multiple $\ppbar$ interactions in the same collision, hadronic jet-energy scale, 
 energy contribution to the event from the fragments of the interacting proton and antiproton (underlying event), 
and out-of-cone energy lost in the energy-clustering procedure.
We vary the corresponding JES parameters~\cite{JetCorrections} by $\pm1$  standard deviation of their estimates 
and build alternative templates for both simulated  signal and background events.
These templates are used to generate  PE's and the average deviations of the results from those obtained
         with default templates
are interpreted as the corresponding systematic uncertainties. 
The individual uncertainties are then summed in quadrature  to obtain the combined JES  uncertainty.

Using the PE's method, we study the systematic uncertainties from sources other than JES  for a few values of $w$.
We estimate these effects by calculating the average deviations between the results of PE's performed with default and modified templates.
The modified templates are derived  by using event samples
generated with variations of the relevant parameters within their uncertainties.
We estimate the modeling uncertainty that stems from the  
difference between leading-order (LO) and next-to-leading-order (NLO)  quantum chromodynamics (QCD) calculations 
by comparing MC samples from LO and NLO generators ({\sc pythia}  and {\sc powheg}). 
The uncertainty arising from the choice of the  MC hadronization model and MC generator is estimated by comparing samples  
generated by using {\sc pythia}  and {\sc herwig}~\cite{herwig}  computer codes. 
The systematic effect due to the lepton-energy scale uncertainty is evaluated by varying the electron energy and muon momentum scales.  
The systematic uncertainty associated with background modeling  accounts  for the variations of the  background template shapes, 
the background composition and the total background  normalization.
The systematic effect due to the imperfect modeling of the initial-state and final-state gluon radiation is estimated 
by varying the {\sc pythia} parameters that control the amount of these radiations. 
To estimate the systematic effect due to the top-quark production mechanism ($gg$ fraction) we vary the relative fractions of 
$q\bar{q} \rightarrow \ttbar$ and $gg \rightarrow \ttbar$ sub-processes in the {\sc pythia}  model by reweighting the gluon  fraction from 5\% to 20\%.
We take into account the additional uncertainty on the $b$-jet-energy scale 
%that occurs 
due to the difference in calorimeter response to jets from light quarks and $b$ quarks 
and the imperfect modeling of the $b$-quark fragmentation and $b$-hadron decay branching fractions.
The  systematic effect due to the difference in the luminosity profile between data and MC is estimated. 
The color reconnection (CR) systematic uncertainty~\cite{Colorreconnection}  is evaluated by comparing {\sc pythia} 
MC samples generated with and without CR effects.
We take into account the systematic effect stemming from the limited size of the MC samples.
We estimate the systematic uncertainty due to parton distribution functions (PDFs) 
by comparing results from two different PDF families, varying the QCD scale, and propagating the uncertainties 
arising from the global fit of the CTEQ6M~\cite{cteq6m} functions.
The systematic uncertainty related to the modeling of the $b$-tagging efficiency is also estimated. 
Details of the systematic uncertainty estimation are in Ref.~\cite{tev_comb}. 

\begin{figure}[htbp]
\begin{center}
\epsfig{file=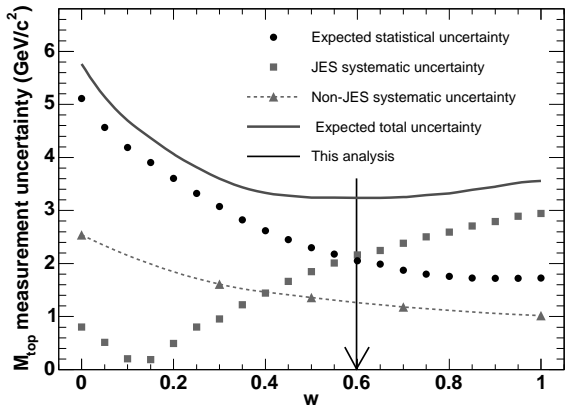 ,scale=0.42}
\end{center}
\caption{
Uncertainties in the measurement of $\Mtop$ as a function of $w$. The  arrow at  $w = 0.6$
shows the minimum of the   expected total uncertainty.
}
\label{_optim_}
\end{figure}

The combined systematic uncertainty generated by sources other than JES (``non-JES uncertainty'') is calculated as
the sum in quadrature of these  uncertainties.
To estimate the non-JES systematic uncertainty for any value of $w$, we use cubic spline interpolations.
The obtained values of the expected statistical, JES, non-JES, and total uncertainties are shown 
as  functions of $w$ in Fig.~\ref{_optim_}. The expected statistical and  JES uncertainties are  changing in the  opposite direction  
as $w$ varies between  0 and 1 
while  the non-JES uncertainty shows a slow falling dependence.
The expected total uncertainty is estimated as the sum in quadrature of the statistical, JES and non-JES  uncertainties and has a 
minimum in the interval between 0.5 and 0.7. 

For  the  data fit, we use  $w$ = 0.6. 
%This value is a center of the  interval where the expected total uncertainty has the minimum.  
We observe a 9\% improvement in the total uncertainty in the case  of $w = 0.6$  with respect to using only 
the reconstructed $\MTreco$ analysis ($w = 1$).

Although $\MTalt$ does not depend explicitly on JES, the $\Mtop$ measurement using only $\MTalt$ 
(points with $w=0$ in Fig.~\ref{_optim_}) is still affected by the JES uncertainty because the JES impacts
the event selection. 
When varying the JES, the change in event sample accepted by
      the selection criteria on variables that depend on jet energies
generates a  change in the $\MTalt$ distribution that affects the $\Mtop$ measurement.
We find that by varying the JES, opposite systematic shifts are induced in  $\MTreco$
and $\MTalt$.
%and the resulting shift for the $\Mtop$ measurement depends on the  $w$ value.  
%using $\MTalt$ and $\MTreco$ 
%(points with $w=0$ and with $w=1$ in Fig.~\ref{_optim_}).
%There are two effects generated by the JES uncertainty:
%an impact on the shape of  variables  being used and the effect on the number of selected events. 
These  systematic  shifts bias  the $\Mtop$ measurement in opposite directions
minimizing the  JES uncertainty  at  $w$=0.12.  This minimum depends only on the  
variables choice, $\MTreco$ and $\MTalt$, and their sensitivity to  $\Mtop$. 
If the sample size would be large such that the statistical uncertainty could be neglected, 
the $w$=0.12 choice would be optimal for this analysis.
%In the case of a high-statistics data sample, $w$=0.12 is the optimal value for this  analysis.  
In that scenario, the JES uncertainty would approximate zero and the non-JES uncertainty would remain as the major contributor to  the total measurement uncertainty.

%The dependence  of the JES systematic uncertainty on $w$ can be understood as follows:
%at large $w$ values the first effect dominates,
%we have a compensation on the interval of the $w$-values from 0.10 to 0.15, and  
%below approximately $w = 0.10$ the second effect starts to dominate.

%~\\\\
\section{\label{sec:Result} Result}
\begin{figure}[htbp]
\begin{center}
\epsfig{file=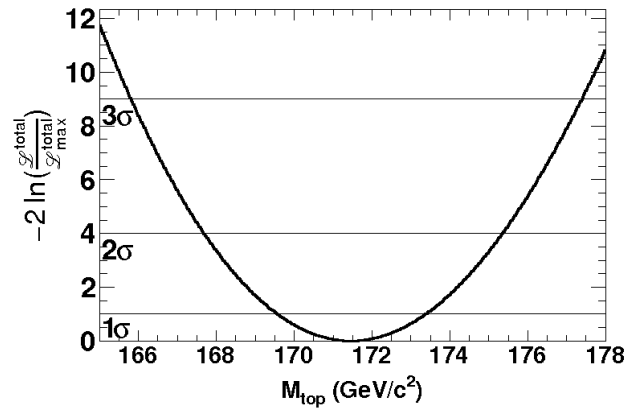,scale=0.42}
\end{center}
\caption{Observed shape of  $-2 \ln(\frac{ \mathscr{L}^{\rm total}}{\mathscr{L}^{\rm total}_{\rm max}})$  as a function of the top-quark mass. Horizontal lines show the 
values corresponding to one, two, and three standard-deviation  uncertainties.
}
\label{lh_data}
\end{figure}

With $w = 0.6$ the fit to the data yields  $\Mtop$ = 171.5~$\pm$~1.9~GeV/$c^2$  including statistical uncertainties only.
The normalized negative log-likelihood function versus the top-quark mass is shown in Fig.~\ref{lh_data}.
Its shape approximates a parabola and the horizontal lines show the values of 
likelihood ratios corresponding to one, two, and three standard-deviation ($\sigma$) uncertainties.
%the  1-$\sigma$,  2-$\sigma$ and 3-$\sigma$  uncertainties 
%assuming Gaussian behavior in the measurement. 

The individual systematic  uncertainties affecting the $\Mtop$ measurement are listed in Table~\ref{systerr_total}. 
The total systematic uncertainty, obtained by adding individual components in quadrature, is 2.5~GeV/$c^2$.
The statistical and total systematic uncertainties combined in quadrature amount to a total uncertainty of 3.2~GeV/$c^2$.

\begin{figure}[htbp]
\begin{center}
\epsfig{file=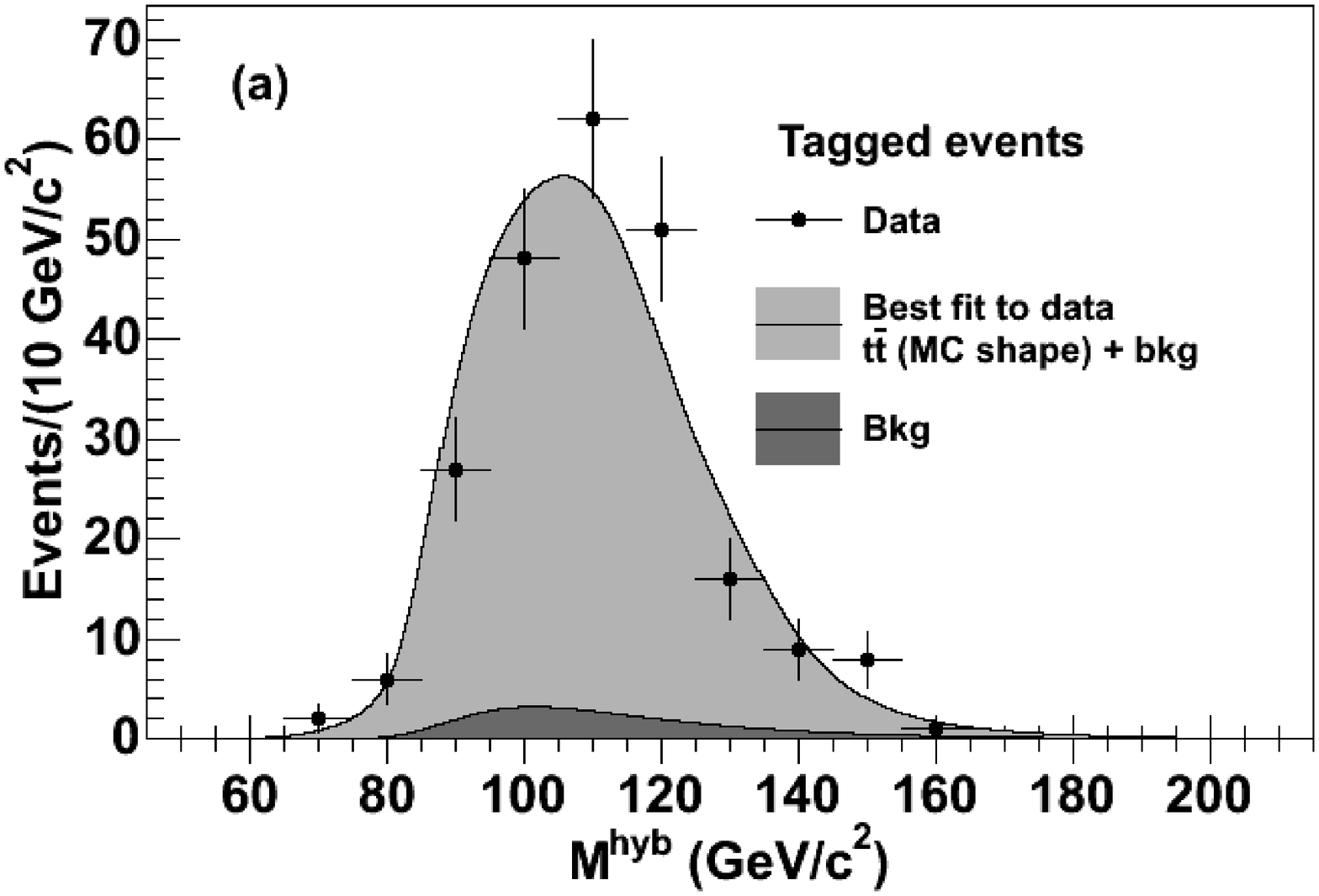,scale=0.34}
\epsfig{file=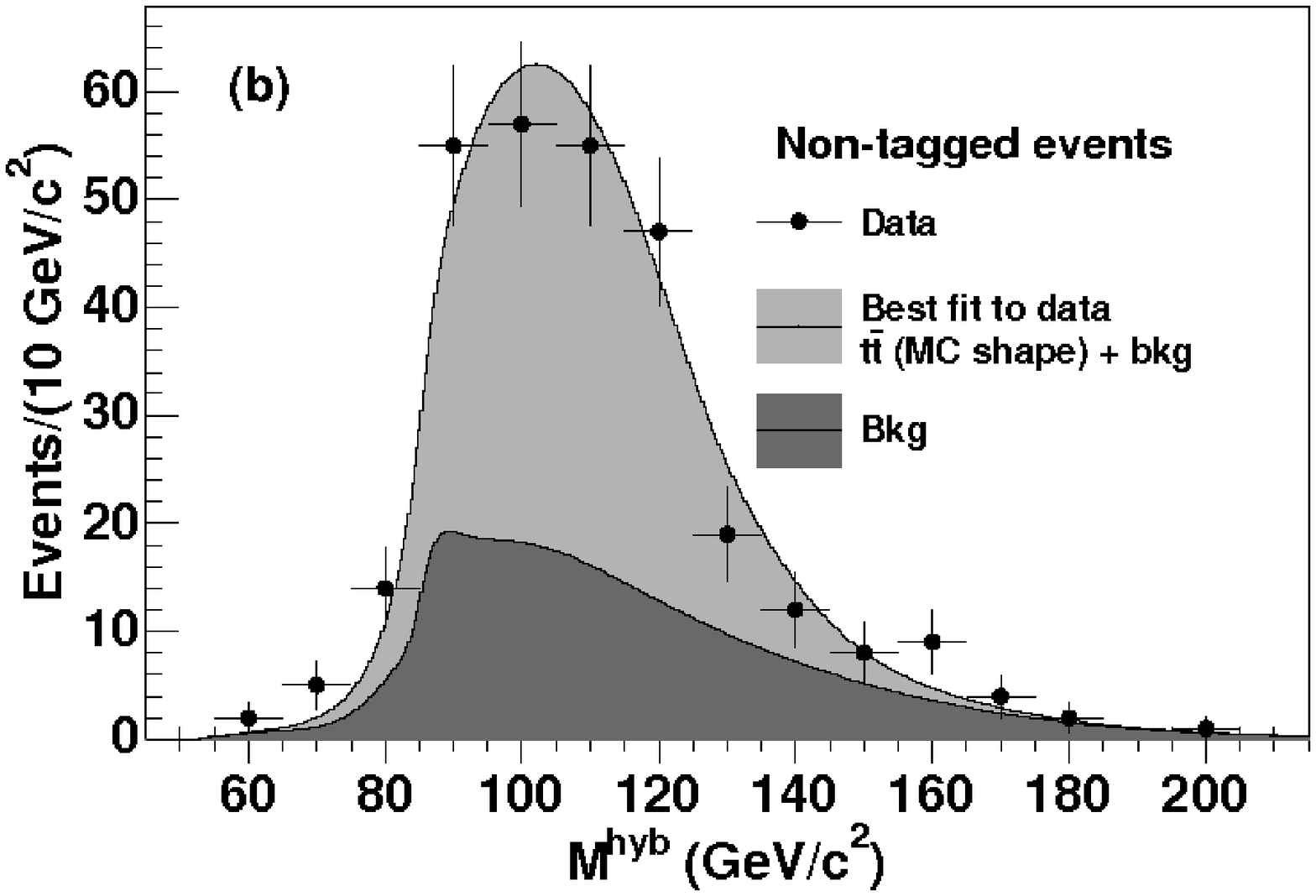,scale=0.42}
\end{center}
\caption{Distribution of the reconstructed variable $\MTeff$.
%Results of applying the likelihood procedure to the data.
The figure shows 
the data (points), the background (dark gray) and  signal (at measured $\Mtop$) plus background (light gray) p.d.f.'s,
normalized accordingly to the fit result. Plots $(a)$ and $(b)$ correspond to $b$-tagged and non-tagged subsamples.
}
\label{Constrained Likelihood Fit}
\end{figure}

Figure~\ref{Constrained Likelihood Fit} shows the $\MTeff$ distributions for $b$-tagged and non-tagged events.
%In Fig.~\ref{Constrained Likelihood Fit}, 
We superimpose the data points to the expected signal  
and background distributions normalized to the numbers of events returned by the fit. 
The signal  distribution  corresponds  to the measured value of $\Mtop$.

\begin{figure}[htbp]
\begin{center}
\epsfig{file=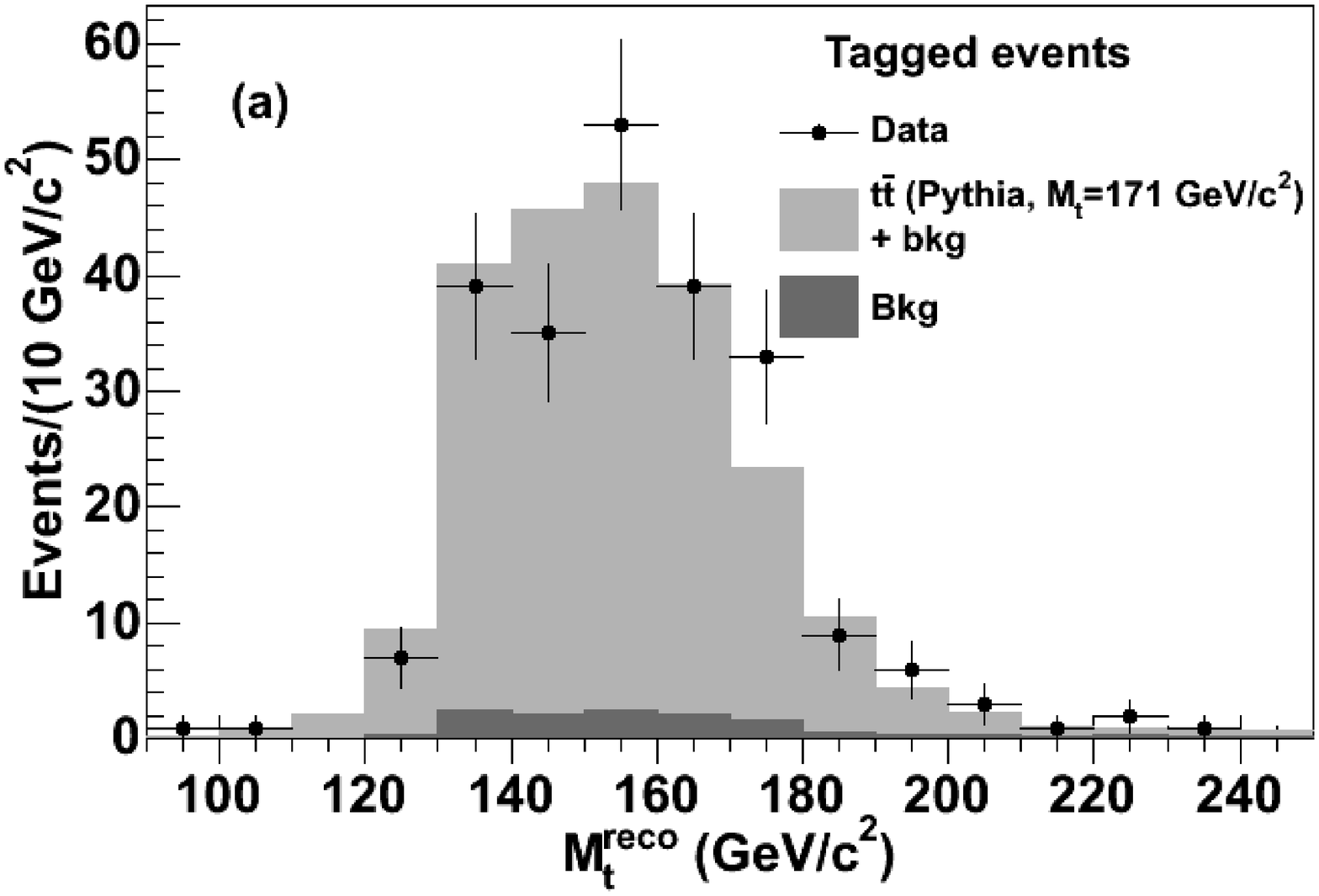,scale=0.34}
\epsfig{file=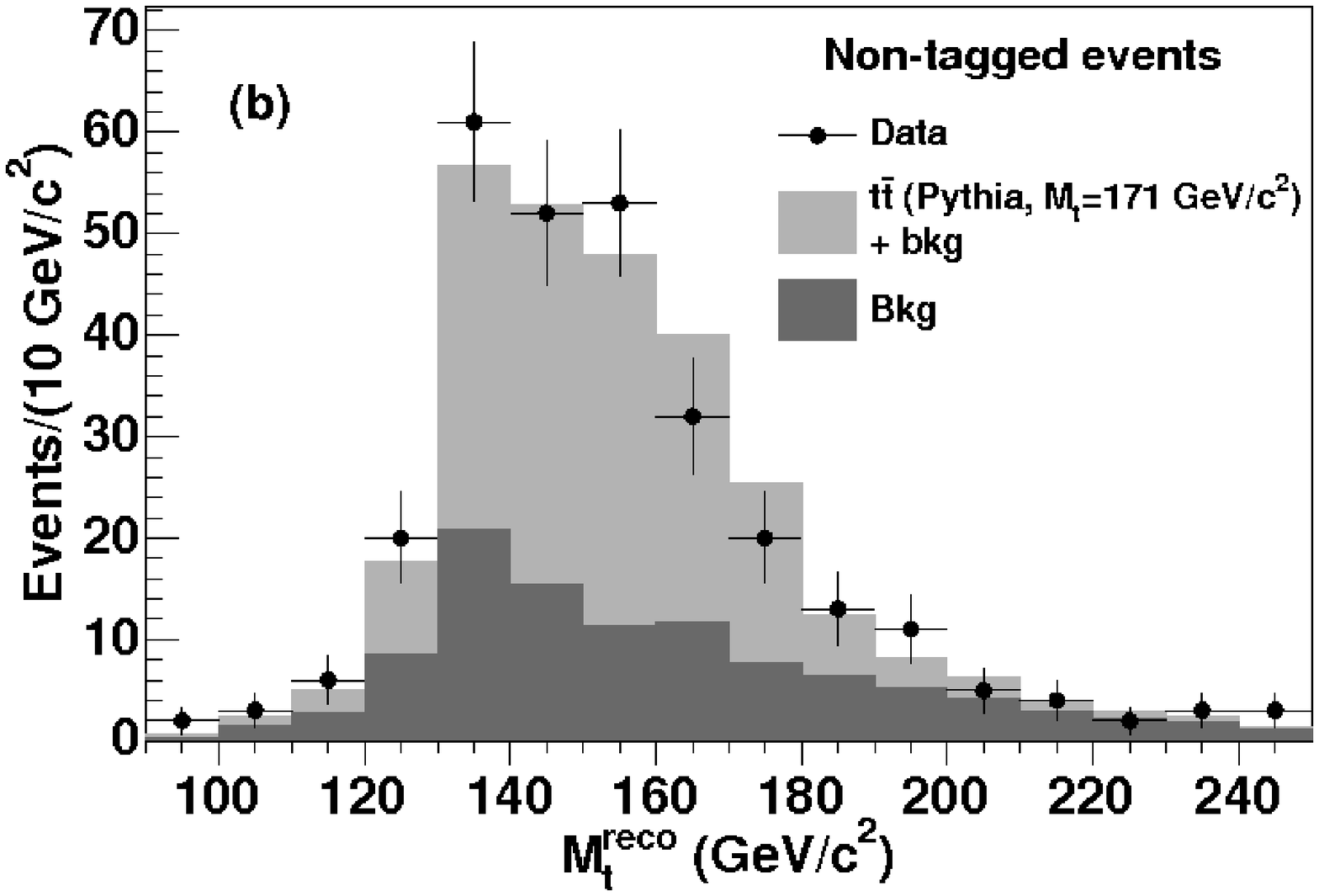,scale=0.42}
\end{center}
\caption{Distributions of reconstructed mass  $\MTreco$  overlaid with the  background
     (dark gray) and signal plus background  (light gray) histograms  in the
     $(a)$ tagged and $(b)$ untagged samples. 
%The top-quark mass value of 171~GeV/$c^2$, closest to the value returned by the data fit, is used for the signal histogram. 
}
\label{mrec_fit}
\end{figure}

\begin{figure}[htbp]
\begin{center}
\epsfig{file=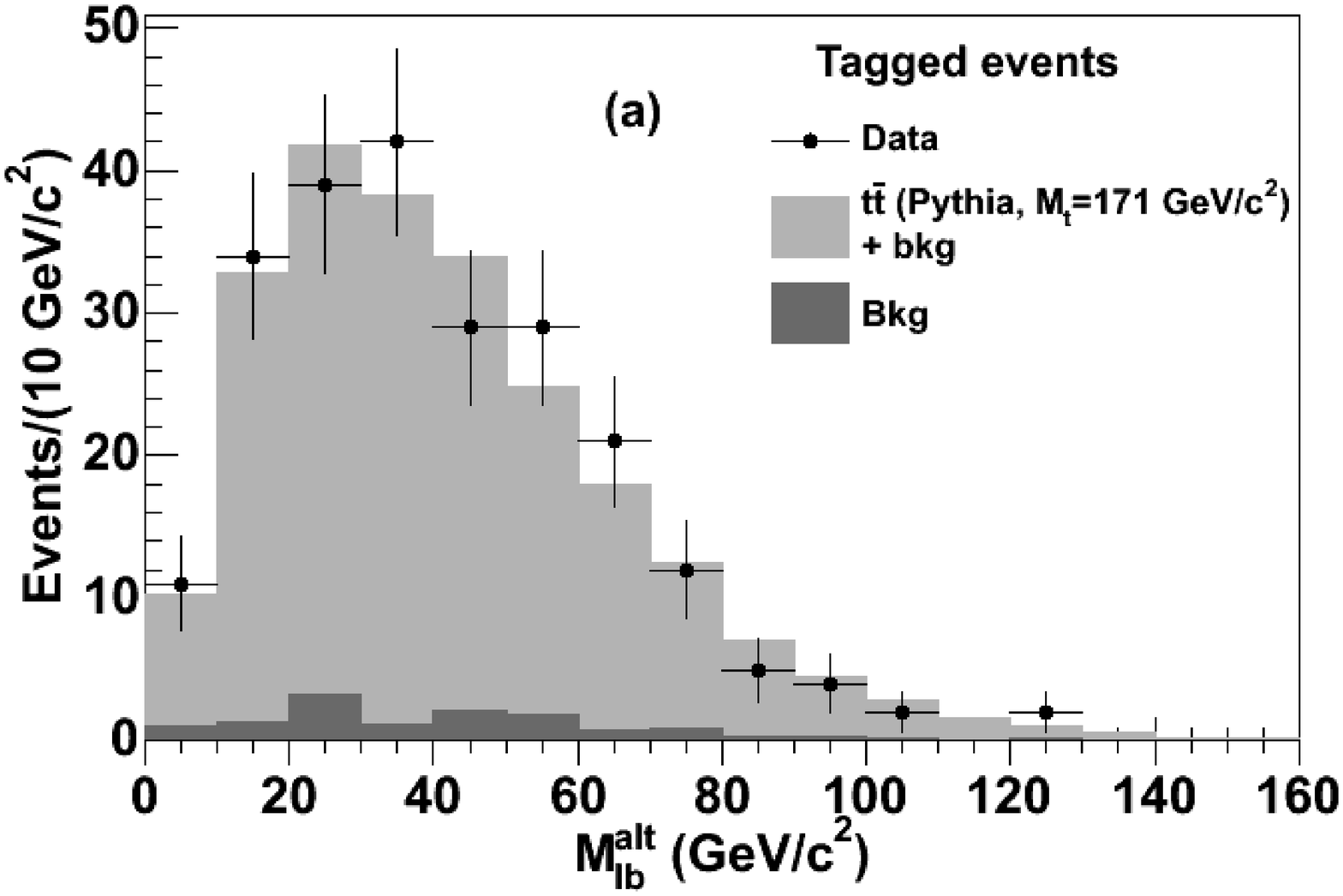,scale=0.34}
\epsfig{file=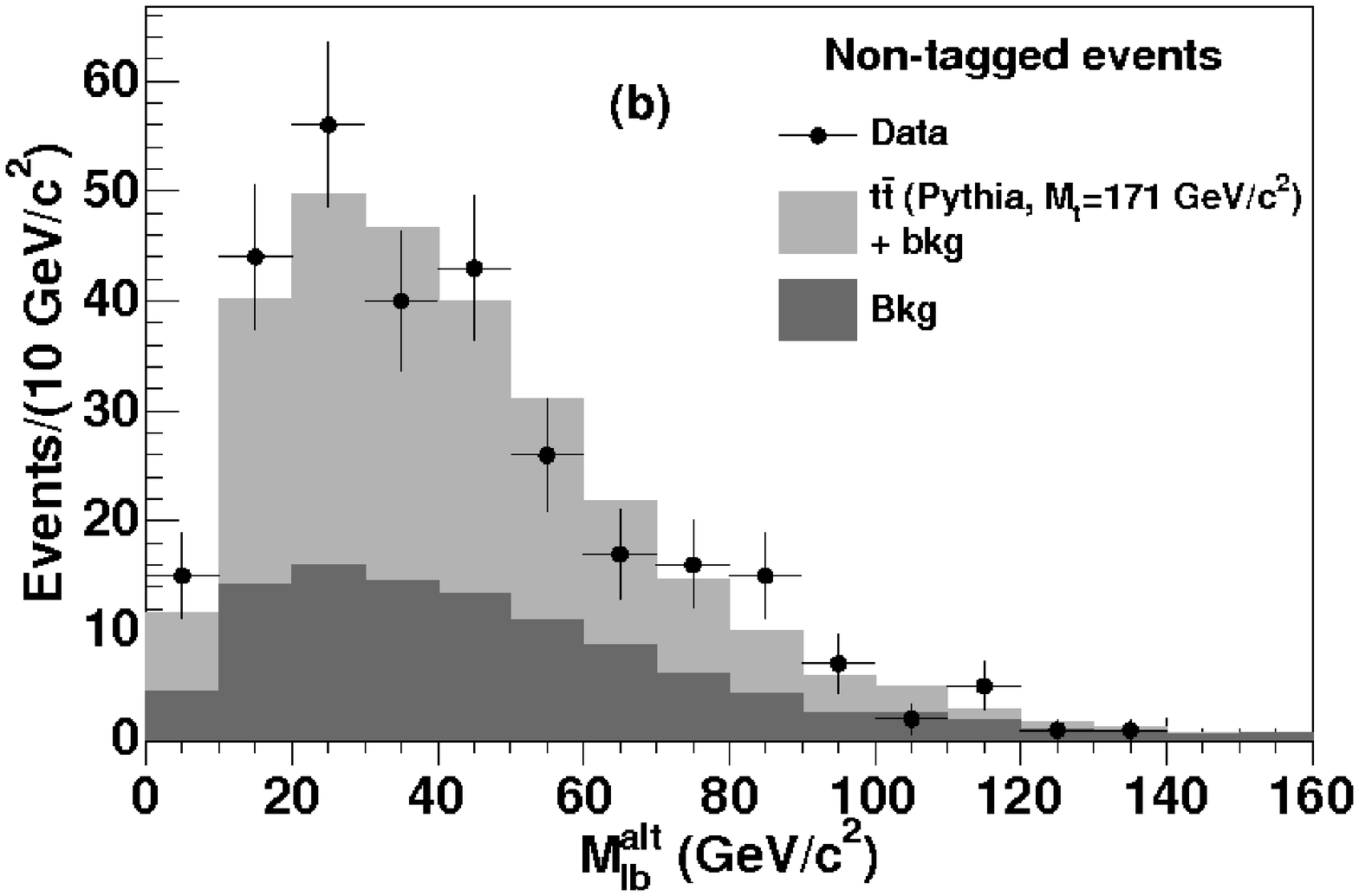,scale=0.42}
\end{center}
\caption{Distributions of reconstructed mass  $\MTalt$  overlaid with  the background
     (dark gray) and signal plus background  (light gray) histograms  in the
     $(a)$ tagged and $(b)$ untagged samples. 
%The top-quark mass value of 171~GeV/$c^2$, closest to the value returned by the data fit,is used for the signal histogram. 
}
\label{var2_fit}
\end{figure}

Similar  plots for the  variables $\MTreco$ and  $\MTalt$ are shown in Figs.~\ref{mrec_fit} and \ref{var2_fit}.
All plots  are normalized  to the numbers of events  returned by the fit. The top-quark mass value of 171~GeV/$c^2$, 
closest to the value returned by the data fit, is used for the signal histogram. 
The p-values for the $\MTreco$ distributions are 71\% and 91\% for the $b$-tagged and non-tagged subsamples.  
For the  $\MTalt$ distributions, the p-values  are  96\% and  55\%  for the $b$-tagged and non-tagged subsamples.   
An excellent  agreement between data and the  simulated distributions is observed. 
%Fig 4(a) p-value=0.71 
%Fig 4(8) p-value=0.91
%Fig 5(a) p-value=0.96
%Fig 5(8) p-value=0.55 

%Figure 2
%Figure 3
%Figure  4
%Figure 5

\begin{table}[htbp]
\caption{Summary of uncertainties.~~~~~~~~~~~~~~~~~~~~~~~~~~~~}
% on the top-quark mass measurement.} 
\label{systerr_total}
\begin{center}
\vspace{0.05in}
\begin{tabular}{lc}
 Source  & Uncertainty (GeV/$c^2$) \\
\cline{1-2}
Jet-energy scale  & 2.2 \\
NLO effects & 0.7 \\
Monte Carlo generators & 0.5 \\
Lepton-energy scale    &    0.4   \\
Background modeling & 0.4 \\
Initial- and final-state radiation & 0.4 \\
$gg$ fraction & 0.3 \\
$b$-jet-energy scale  & 0.3   \\
Luminosity profile & 0.3  \\
Color reconnection & 0.2 \\
MC sample size & 0.2 \\
Parton distribution functions & 0.2  \\
$b$-tagging & 0.1 \\
\cline{1-2} 
Total systematic uncertainty& 2.5 \\
Statistical uncertainty& 1.9 \\
\cline{1-2} 
Total  & 3.2 \\
%\cline{1-2}
\hline
\hline
\end{tabular}

\end{center}
\end{table}

%~\\
\section{\label{sec:Conclusion} Summary}
In conclusion, we  present a measurement of the top-quark mass with $\ttbar$ dilepton events using the full CDF Run~II data set,
which corresponds to an integrated luminosity of 9.1~fb$^{-1}$ from 1.96~TeV $\ppbar$ collisions.
The result  is
$\Mtop = 171.5\pm 1.9~\textrm{(stat)}\pm 2.5~\textrm{(syst)}\ \mbox{GeV}/c^2$. 
The measured value of $\Mtop$ is compatible with the world-average top-quark mass of 
$\Mtop = 173.34 \pm 0.76\ \mbox{GeV}/c^2$~\cite{mtop_comb}. 
This measurement is  the final CDF Run~II result in the dilepton channel  and supersedes the 
previous published value of $\Mtop = 170.3\pm 2.0~\textrm{(stat)}\pm 3.1~\textrm{(syst)}\ \mbox{GeV}/c^2$
~\cite{Hyun-Su5.6}.  
The accuracy achieved is approximately 14\% better than in the previous measurement. 
Most of this improvement, 9\%,  is due to using a new technique for optimizing the 
combined statistical and systematic uncertainty while the rest, 5\%, is due to using a 
larger data sample.
This technique is applicable to a wide range of measurements whose precisions are dominated by systematic uncertainties, 
in which  an optimization between statistical and systematic uncertainty is  required. 
%Specifically for  the  dilepton top-quark measurements, it would allow a significant improvement of 
%the total uncertainty due to the fact that the major systematic contribution decreases with increase of the data sample.

\section{Acknowledgments}
We thank the Fermilab staff and the technical staffs of the
participating institutions for their vital contributions. This work
was supported by the U.S. Department of Energy and National Science
Foundation; the Italian Istituto Nazionale di Fisica Nucleare; the
Ministry of Education, Culture, Sports, Science and Technology of
Japan; the Natural Sciences and Engineering Research Council of
Canada; the National Science Council of the Republic of China; the
Swiss National Science Foundation; the A.P. Sloan Foundation; the
Bundesministerium f\"ur Bildung und Forschung, Germany; the Korean
World Class University Program, the National Research Foundation of
Korea; the Science and Technology Facilities Council and the Royal
Society, United Kingdom; the Russian Foundation for Basic Research;
the Ministerio de Ciencia e Innovaci\'{o}n, and Programa
Consolider-Ingenio 2010, Spain; the Slovak R\&D Agency; the Academy
of Finland; the Australian Research Council (ARC); and the EU community
Marie Curie Fellowship Contract No. 302103.

\bibliography{bibliography13}

\end{document}

%% file: author.tex
% Last update: $Date: 2015/04/07 19:40:02 $
\affiliation{Institute of Physics, Academia Sinica, Taipei, Taiwan 11529, Republic of China}
\affiliation{Argonne National Laboratory, Argonne, Illinois 60439, USA}
\affiliation{University of Athens, 157 71 Athens, Greece}
\affiliation{Institut de Fisica d'Altes Energies, ICREA, Universitat Autonoma de Barcelona, E-08193, Bellaterra (Barcelona), Spain}
\affiliation{Baylor University, Waco, Texas 76798, USA}
\affiliation{Istituto Nazionale di Fisica Nucleare Bologna, \ensuremath{^{kk}}University of Bologna, I-40127 Bologna, Italy}
\affiliation{University of California, Davis, Davis, California 95616, USA}
\affiliation{University of California, Los Angeles, Los Angeles, California 90024, USA}
\affiliation{Instituto de Fisica de Cantabria, CSIC-University of Cantabria, 39005 Santander, Spain}
\affiliation{Carnegie Mellon University, Pittsburgh, Pennsylvania 15213, USA}
\affiliation{Enrico Fermi Institute, University of Chicago, Chicago, Illinois 60637, USA}
\affiliation{Comenius University, 842 48 Bratislava, Slovakia; Institute of Experimental Physics, 040 01 Kosice, Slovakia}
\affiliation{Joint Institute for Nuclear Research, RU-141980 Dubna, Russia}
\affiliation{Duke University, Durham, North Carolina 27708, USA}
\affiliation{Fermi National Accelerator Laboratory, Batavia, Illinois 60510, USA}
\affiliation{University of Florida, Gainesville, Florida 32611, USA}
\affiliation{Laboratori Nazionali di Frascati, Istituto Nazionale di Fisica Nucleare, I-00044 Frascati, Italy}
\affiliation{University of Geneva, CH-1211 Geneva 4, Switzerland}
\affiliation{Glasgow University, Glasgow G12 8QQ, United Kingdom}
\affiliation{Harvard University, Cambridge, Massachusetts 02138, USA}
\affiliation{Division of High Energy Physics, Department of Physics, University of Helsinki, FIN-00014, Helsinki, Finland; Helsinki Institute of Physics, FIN-00014, Helsinki, Finland}
\affiliation{University of Illinois, Urbana, Illinois 61801, USA}
\affiliation{The Johns Hopkins University, Baltimore, Maryland 21218, USA}
\affiliation{Institut f\"{u}r Experimentelle Kernphysik, Karlsruhe Institute of Technology, D-76131 Karlsruhe, Germany}
\affiliation{Center for High Energy Physics: Kyungpook National University, Daegu 702-701, Korea; Seoul National University, Seoul 151-742, Korea; Sungkyunkwan University, Suwon 440-746, Korea; Korea Institute of Science and Technology Information, Daejeon 305-806, Korea; Chonnam National University, Gwangju 500-757, Korea; Chonbuk National University, Jeonju 561-756, Korea; Ewha Womans University, Seoul, 120-750, Korea}
\affiliation{Ernest Orlando Lawrence Berkeley National Laboratory, Berkeley, California 94720, USA}
\affiliation{University of Liverpool, Liverpool L69 7ZE, United Kingdom}
\affiliation{University College London, London WC1E 6BT, United Kingdom}
\affiliation{Centro de Investigaciones Energeticas Medioambientales y Tecnologicas, E-28040 Madrid, Spain}
\affiliation{Massachusetts Institute of Technology, Cambridge, Massachusetts 02139, USA}
\affiliation{University of Michigan, Ann Arbor, Michigan 48109, USA}
\affiliation{Michigan State University, East Lansing, Michigan 48824, USA}
\affiliation{Institution for Theoretical and Experimental Physics, ITEP, Moscow 117259, Russia}
\affiliation{University of New Mexico, Albuquerque, New Mexico 87131, USA}
\affiliation{The Ohio State University, Columbus, Ohio 43210, USA}
\affiliation{Okayama University, Okayama 700-8530, Japan}
\affiliation{Osaka City University, Osaka 558-8585, Japan}
\affiliation{University of Oxford, Oxford OX1 3RH, United Kingdom}
\affiliation{Istituto Nazionale di Fisica Nucleare, Sezione di Padova, \ensuremath{^{ll}}University of Padova, I-35131 Padova, Italy}
\affiliation{University of Pennsylvania, Philadelphia, Pennsylvania 19104, USA}
\affiliation{Istituto Nazionale di Fisica Nucleare Pisa, \ensuremath{^{mm}}University of Pisa, \ensuremath{^{nn}}University of Siena, \ensuremath{^{oo}}Scuola Normale Superiore, I-56127 Pisa, Italy, \ensuremath{^{pp}}INFN Pavia, I-27100 Pavia, Italy, \ensuremath{^{qq}}University of Pavia, I-27100 Pavia, Italy}
\affiliation{University of Pittsburgh, Pittsburgh, Pennsylvania 15260, USA}
\affiliation{Purdue University, West Lafayette, Indiana 47907, USA}
\affiliation{University of Rochester, Rochester, New York 14627, USA}
\affiliation{The Rockefeller University, New York, New York 10065, USA}
\affiliation{Istituto Nazionale di Fisica Nucleare, Sezione di Roma 1, \ensuremath{^{rr}}Sapienza Universit\`{a} di Roma, I-00185 Roma, Italy}
\affiliation{Mitchell Institute for Fundamental Physics and Astronomy, Texas A\&M University, College Station, Texas 77843, USA}
\affiliation{Istituto Nazionale di Fisica Nucleare Trieste, \ensuremath{^{ss}}Gruppo Collegato di Udine, \ensuremath{^{tt}}University of Udine, I-33100 Udine, Italy, \ensuremath{^{uu}}University of Trieste, I-34127 Trieste, Italy}
\affiliation{University of Tsukuba, Tsukuba, Ibaraki 305, Japan}
\affiliation{Tufts University, Medford, Massachusetts 02155, USA}
\affiliation{University of Virginia, Charlottesville, Virginia 22906, USA}
\affiliation{Waseda University, Tokyo 169, Japan}
\affiliation{Wayne State University, Detroit, Michigan 48201, USA}
\affiliation{University of Wisconsin, Madison, Wisconsin 53706, USA}
\affiliation{Yale University, New Haven, Connecticut 06520, USA}

\author{T.~Aaltonen}
\affiliation{Division of High Energy Physics, Department of Physics, University of Helsinki, FIN-00014, Helsinki, Finland; Helsinki Institute of Physics, FIN-00014, Helsinki, Finland}
\author{S.~Amerio\ensuremath{^{ll}}}
\affiliation{Istituto Nazionale di Fisica Nucleare, Sezione di Padova, \ensuremath{^{ll}}University of Padova, I-35131 Padova, Italy}
\author{D.~Amidei}
\affiliation{University of Michigan, Ann Arbor, Michigan 48109, USA}
\author{A.~Anastassov\ensuremath{^{w}}}
\affiliation{Fermi National Accelerator Laboratory, Batavia, Illinois 60510, USA}
\author{A.~Annovi}
\affiliation{Laboratori Nazionali di Frascati, Istituto Nazionale di Fisica Nucleare, I-00044 Frascati, Italy}
\author{J.~Antos}
\affiliation{Comenius University, 842 48 Bratislava, Slovakia; Institute of Experimental Physics, 040 01 Kosice, Slovakia}
\author{G.~Apollinari}
\affiliation{Fermi National Accelerator Laboratory, Batavia, Illinois 60510, USA}
\author{J.A.~Appel}
\affiliation{Fermi National Accelerator Laboratory, Batavia, Illinois 60510, USA}
\author{T.~Arisawa}
\affiliation{Waseda University, Tokyo 169, Japan}
\author{A.~Artikov}
\affiliation{Joint Institute for Nuclear Research, RU-141980 Dubna, Russia}
% ...

\collaboration{CDF Collaboration}
\altaffiliation[With visitors from]{
\ensuremath{^{a}}University of British Columbia, Vancouver, BC V6T 1Z1, Canada,
\ensuremath{^{b}}Istituto Nazionale di Fisica Nucleare, Sezione di Cagliari, 09042 Monserrato (Cagliari), Italy,
\ensuremath{^{c}}University of California Irvine, Irvine, CA 92697, USA,
\ensuremath{^{d}}Institute of Physics, Academy of Sciences of the Czech Republic, 182~21, Czech Republic,
\ensuremath{^{e}}CERN, CH-1211 Geneva, Switzerland,
\ensuremath{^{f}}Cornell University, Ithaca, NY 14853, USA,
\ensuremath{^{g}}University of Cyprus, Nicosia CY-1678, Cyprus,
\ensuremath{^{h}}Office of Science, U.S. Department of Energy, Washington, DC 20585, USA,
\ensuremath{^{i}}University College Dublin, Dublin 4, Ireland,
\ensuremath{^{j}}ETH, 8092 Z\"{u}rich, Switzerland,
\ensuremath{^{k}}University of Fukui, Fukui City, Fukui Prefecture, Japan 910-0017,
\ensuremath{^{l}}Universidad Iberoamericana, Lomas de Santa Fe, M\'{e}xico, C.P. 01219, Distrito Federal,
\ensuremath{^{m}}University of Iowa, Iowa City, IA 52242, USA,
\ensuremath{^{n}}Kinki University, Higashi-Osaka City, Japan 577-8502,
\ensuremath{^{o}}Kansas State University, Manhattan, KS 66506, USA,
\ensuremath{^{p}}Brookhaven National Laboratory, Upton, NY 11973, USA,
\ensuremath{^{q}}Istituto Nazionale di Fisica Nucleare, Sezione di Lecce, Via Arnesano, I-73100 Lecce, Italy,
\ensuremath{^{r}}Queen Mary, University of London, London, E1 4NS, United Kingdom,
\ensuremath{^{s}}University of Melbourne, Victoria 3010, Australia,
\ensuremath{^{t}}Muons, Inc., Batavia, IL 60510, USA,
\ensuremath{^{u}}Nagasaki Institute of Applied Science, Nagasaki 851-0193, Japan,
\ensuremath{^{v}}National Research Nuclear University, Moscow 115409, Russia,
\ensuremath{^{w}}Northwestern University, Evanston, IL 60208, USA,
\ensuremath{^{x}}University of Notre Dame, Notre Dame, IN 46556, USA,
\ensuremath{^{y}}Universidad de Oviedo, E-33007 Oviedo, Spain,
\ensuremath{^{z}}CNRS-IN2P3, Paris, F-75205 France,
\ensuremath{^{aa}}Universidad Tecnica Federico Santa Maria, 110v Valparaiso, Chile,
\ensuremath{^{bb}}Sejong University, Seoul, South Korea,
\ensuremath{^{cc}}The University of Jordan, Amman 11942, Jordan,
\ensuremath{^{dd}}Universite catholique de Louvain, 1348 Louvain-La-Neuve, Belgium,
\ensuremath{^{ee}}University of Z\"{u}rich, 8006 Z\"{u}rich, Switzerland,
\ensuremath{^{ff}}Massachusetts General Hospital, Boston, MA 02114 USA,
\ensuremath{^{gg}}Harvard Medical School, Boston, MA 02114 USA,
\ensuremath{^{hh}}Hampton University, Hampton, VA 23668, USA,
\ensuremath{^{ii}}Los Alamos National Laboratory, Los Alamos, NM 87544, USA,
\ensuremath{^{jj}}Universit\`{a} degli Studi di Napoli Federico I, I-80138 Napoli, Italy
}
\noaffiliation
% Last update: $Date: 2015/04/07 19:40:02 $

%% file: dilmtop_prd_v14.bbl
\begin{thebibliography}{34}
\bibitem[{csy()}]{csyst}
\bibinfo{note}{{A cylindrical ($r$, $\phi$, $z$) coordinate system is used. The
  origin of the reference system is the geometric center of the detector, with
  the $ z$ axis pointing along the proton beam. The pseudorapidity $\eta$ is
  defined by $\eta = -\ln (\tan (\theta/2))$, where $\theta$ is the polar angle
  relative to the $z$ axis. The transverse momentum and energy of a detected
  particle or jet are defined as $\pt = p\sin\theta$ and $\et = E\sin\theta$,
  respectively, where $p$ and $E$ are the momentum and energy of the particle
  or jet.}}
\bibitem[{iso()}]{isol}
\bibinfo{note}{{A lepton is isolated when the transverse energy measured in the
  calorimeter not associated with the lepton and calculated in a cone of radius
  $\Delta R=\sqrt{\Delta \eta^2+\Delta \phi^2}=0.4$ about the lepton trajectory
  does not exceed 10$\%$ of the lepton $E_T$ ($\pt$ for muons).}}
\bibitem[{met({\natexlab{a}})}]{met_def}
\bibinfo{note}{{The missing transverse energy, an imbalance of energy in the
  transverse plane of the detector, is defined by $\met=|-\sum_i
  E_T^i\hat{n}_i|$ where the sum is performed over the calorimeter towers. The
  quantity $\hat{n}_i$ is the unit vector normal to the beam and pointing from
  the interaction vertex to the $i$-th tower. The $\met$ for events with
  identified muons is corrected by the measured muon transverse momentum.}}
\bibitem[{met({\natexlab{b}})}]{met_sig}
\bibinfo{note}{{The variable named $\met$ significance is defined as
  $\met/\sqrt{E_T^{\rm sum}}$. The value of $E_T^{\rm sum}$ is calculated as
  the sum of transverse energies deposited in calorimeter towers and is
  corrected for the muon transverse momenta, if muons are identifed, and
  various other instrumental effects.}}
\bibitem[{ht_()}]{ht_def}
\bibinfo{note}{{$H_T$ is defined as the transverse-energy scalar sum over
  leptons, jets, and $\met$.}}
\end{thebibliography}
